# In Situ Disdrometer Calibration Using Multiple DSD Moments

John E. LANE[1], Takis KASPARIS[2],
Philip T. METZGER[3], and W. Linwood JONES[4]

[1]Easi-ESC, GMRO Lab, Kennedy Space Center, FL, USA
john.e.lane@nasa.gov

[2]Cyprus University of Technology, Lemesos, Cyprus
kasparis@ucf.edu

[3]NASA Granular Mechanics and Regolith Operations, KSC, FL, USA
philip.t.metzger@nasa.gov

[4] CFRSL, University of Central Florida, Orlando, FL, USA
ljones@ucf.edu

In situ calibration is a proposed strategy for continuous as well as initial calibration of an impact disdrometer. In previous work, a collocated tipping bucket had been utilized to provide a rainfall rate based ~11/3 moment reference to an impact disdrometer's signal processing system for implementation of adaptive calibration. Using rainfall rate only, transformation of impulse amplitude to a drop volume based on a simple power law was used to define an error surface in the model's parameter space. By incorporating optical extinction second moment measurements with rainfall rate data, an improved in situ disdrometer calibration algorithm results due to utilization of multiple (two or more) independent moments of the drop size distribution in the error function definition. The resulting improvement in calibration performance can be quantified by detailed examination of the parameter space error surface using simulation as well as real data.

## 1. INTRODUCTION

Instruments that measure a property of rainfall, often measure a *moment* of the drop size distribution (DSD), expressed by $N(D)$ [$m^{-3}\cdot m^{-1}$]. (Note that *si* units are used throughout this paper in order to avoid complications arising from integrating the DSD over quantities containing a mixed set of units. This may lead to awkward numbers at times, but it is usually a simple matter to convert back to standard units for plotting purposes or comparison to familiar values). The DSD moment is defined as:

$$M_n \equiv \int_0^\infty D^n N(D)\, dD \quad . \tag{1}$$

Weather radar measures the sixth moment of the DSD ($n = 6$). A tipping bucket rain gauge measures approximately the 11/3 moment ($n = 3 + 2/3$), where $D^3$ corresponds to equivalent spherical drop volume and $D^{2/3}$ is the Atlas and Ulbrich (1977) drop size dependence of the terminal velocity approximation. Optical extinction of a laser measures the second moment ($n = 2$). A disdrometer measures the DSD flux which is related to the DSD via the drop terminal velocity function. Note that in this paper DSD, $N(D)$, and *drop spectrum* all describe the same physical quantity, the number drops aloft per volume [$m^{-3}$] per drop size [$m^{-1}$]. Disdrometer *drop spectrum flux* is a related quantity represented by $D(t)$ and is the quantity measured by a calibrated disdrometer, displayed as a scatter plot of all individual drop sizes measured versus time of measurement.

An impact disdrometer is typically calibrated by single drops of known size falling at terminal velocity. Terminal velocity for large drops requires a substantial height of fall, at least 10 m or more. A disdrometer calibrated this way may have a very different response to normal and high rainfall rate conditions, which may lead to large measurement errors, analogous to tipping bucket errors under high rainfall rate conditions. One way to solve this problem is to calibrate a disdrometer under real-time conditions, or in situ calibration.

Optical disdrometers based on processing signals generated by single drops passing through a laser are well-known and have been used effectively (Löffler *et al*. 2000). Measuring optical extinction of visible and near visible light has long been recognized as a means to qualitatively characterize rainfall along a path length of meters to kilometers (Atlas 1953, Uijlenhoet *et al*. 2011). An in situ comparison of rain gauges to disdrometers has been used to address questions related to disdrometer measurement uncertainties (Tokay et al. 2013). Spatial variability of the DSD has been carefully studied near Ciudad Real, Spain using 16 laser disdrometers (Tapiador *et al*. 2010). The researchers concluded that additional disdrometers were needed to adequately characterize the details of the DSD's spatial variability and temporal evolution.

A goal of this paper is to describe a method using laser scattering as a DSD second moment observable to supplement rain gauge based rain rate DSD moment observable ~11/3 (or ~7/2 if using a Gunn and Kinzer (1949) terminal velocity approximation). The philosophy of this approach is that other observables may be included when possible. Other observables might include the $n$ = 6 moment from microwave backscatter such as weather radar or preferably, a small short range microwave backscatter system (Prodi *et. al*. 2011). Since the spatial and temporal disparity of weather radar generally prohibits practical use as a means to calibrate a disdrometer, only two DSD moment sources are discussed in detail in the following sections. The mathematical techniques presented can be expanded to include additional DSD moment sources.

The temporal resolution of a disdrometer is limited only by the decay time of the sensor impulse signal, approximately 0.1 to 30 [ms], a function of drop size. The temporal resolution of a tipping bucket is based on the catch bucket size and main opening diameter. This typically leads to a minimum resolution of a few seconds (limited by the mechanical response of the tipping mechanism) for very high rainfall rates, to very long times for trace rainfall rate, which may then be corrupted by evaporation. Laser/camera extinction is temporally limited by the frame rate of the camera, typically 30 fps. Since the tipping bucket is the limit for the inter-comparison of these three instruments, the variable tip time interval is a convenient parameter for synchronizing all measurements.

The circumstances under investigation in this paper exclude the case of disdrometer to disdrometer comparison and single drop calibration. The focus of this paper is on instruments such as the tipping bucket rain gauge that provide a comparison measurement for disdrometer performance verification and/or calibration under in situ conditions of naturally occurring rainfall. Other instruments that provide collocated measurements are laser extinction devices (similar to a runway visual transmissometer) and microwave radar. Radar will not be discussed in this paper since it is well understood and the problems with weather radar reflectivity as a disdrometer verification/calibration are due to the large differences in temporal and spatial sampling. A short range microwave system (similar in principle to a police radar gun) should solve the spatial and temporal disparity problems.

## 2. HAIL DISDROMETERS AND THE 3D-DSD

Hail disdrometers developed at the Kennedy Space Center were operated at shuttle launch pads 39A and 39B from 2006 through the end of the Space Shuttle program in 2011. In situ calibration procedures and a 3D-DSD interpolation/extrapolation model were successfully applied to a number of hail events during the period of operation (Lane et al. 2008). Since three hail disdrometers were deployed in a triangle around the launch pad, interpolation and extrapolation using hydrometeor trajectory dynamics provided a means for the 3D-DSD model to approximate a hail size distribution (HSD) in a 0.5 km (height) and 1.0 $km^2$ (base) surrounding the launch vehicle. By computing the sixth moment of the HSD, a direct comparison was made to the Melbourne radar volume (see Figure 1).

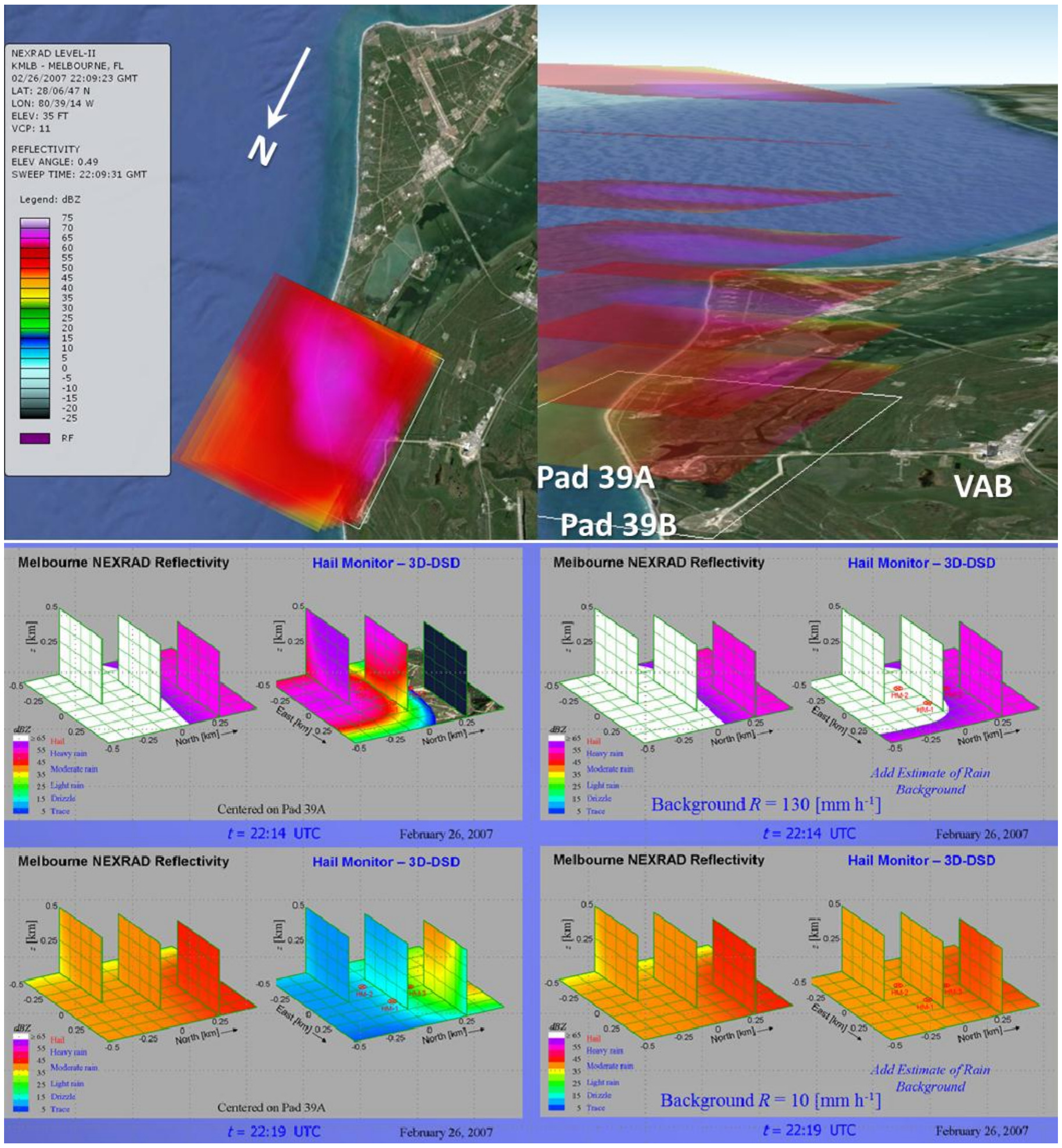

Fig. 1. Damaging hail event during STS-117 processing at Pad 39-A: (top) Melbourne NEXRAD reflectivity; (bottom) 3D-DSD model based spatial and temporal interpolation of hail size distribution based on hail disdrometer array measurements.

By treating the DSD as an unknown distribution function of hydrometeor size $D$, as well as $x$, $y$, $z$, and $t$, all data that measures some moment of the DSD, as well as disdrometer measurements at one or more locations, can provide input to an empirical model, resulting in an approximation of a complete DSD function. This 3D-DSD model must also include an estimate of the vertical and horizontal wind components as a function of $x$, $y$, $z$, and $t$. This is accomplished by using an empirical model of vertical and horizontal wind flow. Estimates of evaporation are also provided to the 3D-DSD model if possible.

Even though hail disdrometers and the 3D-DSD model are not the immediate subject of the paper, they are introduced to provide an example of why it would be useful to deploy a dense network of disdrometers for analysis of the spatial and temporal variability of hydrometeor size distributions.

## 3. DHD FABRICATION

During the 2009-2011 joint project between Cyprus University of Technology (CUT) and University of Central Florida (UCF), numerous iterations of potential low-cost disdrometer prototypes were fabricated and tested. Design goals included use of COTS piezoelectric buzzer disks of various sizes and in various combinations with an electrically isolating moisture barrier encapsulating material. In all iterations, the total sensing area was limited to a size range of 50 to 100 cm$^2$. For reference, the Joss–Waldvogel disdrometer (a meteorological standard) sensor area is 50 cm$^2$. The size options of COTS piezoelectric disks are limited to a few standard diameters. The largest diameter that was found as an available COTS component was the muRata 7NB-41-1 piezoelectric diaphragm, with a ceramic diameter of 25 mm and 41 mm diameter nickel alloy substrate.

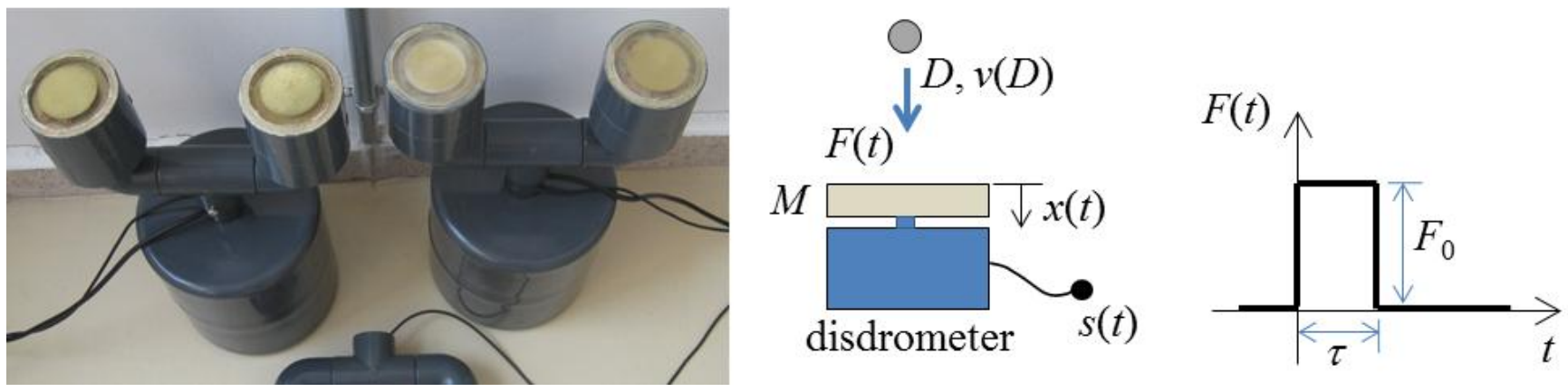

Fig. 2. Left: Two CUT-DHDs using the muRata 7NB-41-1; Middle: SDOF model of disdrometer, where $x(t)$ is the displacement generated by a drop of diameter $D$, and electrical signal $s(t)$ proportional to $x(t)$.; Right: drop impulse force $F(t)$.

The goal of the encapsulating material is to provide a moisture seal, but an equally important purpose is to provide mass loading and damping to the piezoelectric disk. Various encapsulating materials were used from hard marine epoxy with a Shore D hardness of 72 to a soft Cytec Conathane EN-12 polyurethane with a Shore A hardness of 50. Many of the configurations tested consisted of an additional thin plastic cover with a milled angled slope to encourage water roll off. During fabrication, it appeared that fewer bubbles formed in the hard epoxy than in the soft urethane. The best overall solution was to let the encapsulant cure slowly by fine tuning the ratio of part B (hardner) to part A (epoxy). Figure 2 shows the final dual-head configuration, with a total area of 58 $cm^2$. The final dual head configuration is a consequence of utilizing the largest piezoelectric discs commercially available and achieving a practical sensing area in the range of 50 to 100 $cm^2$.

## 4. SDOF MODEL OF IMPACT SENSOR

An impact disdrometer can be approximately modelled as a single degree of freedom (SDOF) system. The goal of the model is to provide some insight into the sensor response, which then helps guide the signal processing design. The SDOF model is diagrammed in Figure 2, where the impulse force is approximated as a square pulse of width $\tau$. The electrical signal $s(t)$ is proportional to the displacement $x(t)$ caused by a drop impact on the sensor surface. The differential equation describing this interaction is:

$$\ddot{x}(t) + \gamma\,\dot{x}(t) + \beta\,x(t) = F(t)/M \quad , \tag{2}$$

where $\gamma$ is a damping coefficient, $\beta = \omega_0^2$ (resonant frequency squared), $F(t)$ is the drop force, and $M$ is the effective mass of the transducer. The solution to Equation (2), using roots of the *characteristic equation*: $\lambda_1 = -\frac{1}{2}\gamma + \frac{1}{2}\sqrt{\gamma^2 - 4\beta}$ , $\lambda_2 = -\frac{1}{2}\gamma - \frac{1}{2}\sqrt{\gamma^2 - 4\beta}$, is:

$$x(t) = \begin{cases} 0 & x < 0 \\ c_1 e^{\lambda_1 t} + c_2 e^{\lambda_2 t} + c_3 & 0 \le x \le \tau \\ d_1 e^{\lambda_1 t} + d_2 e^{\lambda_2 t} & x > \tau \end{cases} \quad . \tag{3}$$

The unknowns in Equation (3) are determined by matching boundary conditions between regions:

$$c_1 = -c_0\left(1+\frac{\gamma}{\sqrt{\gamma^2+4\omega_0^2}}\right) \quad c_2 = -c_0\left(1-\frac{\gamma}{\sqrt{\gamma^2+4\omega_0^2}}\right)$$

$$d_1 = c_0\frac{\left(\gamma+\sqrt{\gamma^2-4\omega_0^2}\right)}{\sqrt{\gamma^2-4\omega_0^2}}\cdot\left(e^{-\frac{1}{2}\tau\left(\gamma-\sqrt{\gamma^2-4\omega_0^2}\right)}-1\right) \tag{4}$$

$$d_2 = c_0\frac{\left(-\gamma+\sqrt{\gamma^2-4\omega_0^2}\right)}{\sqrt{\gamma^2-4\omega_0^2}}\cdot\left(e^{-\frac{1}{2}\tau\left(\gamma+\sqrt{\gamma^2-4\omega_0^2}\right)}-1\right)$$

where $c_0 = F_0/(\omega_0^2 M)$ and $c_3 = c_0$. The impulse shape depends on the drop size *D*:

$$F_0 = \frac{dP}{dt} \approx \frac{m\,v(D)}{\tau}(1+e) \quad , \tag{5}$$

where $m = \pi D^3/6$, $v(D) = \mu D^\chi$, and $\tau = \xi D/v(D)$. The coefficient of restitution *e* is a value between 0 and 1, equal to the relative speed after collision divided by the relative speed before collision. The terminal velocity relation is based on a power law where $\mu$ = 380.7 m s$^{-1}$ m$^{-2/3}$ and $\chi$ = 2/3 (Atlas and Ulbrich 1977). The parameter $\xi$ is an empirical adjustment used to match the sensor response data, typically set to a constant between 0.5 and 1. Figure 3 shows simulated response curves for several drop sizes using Equation (3).

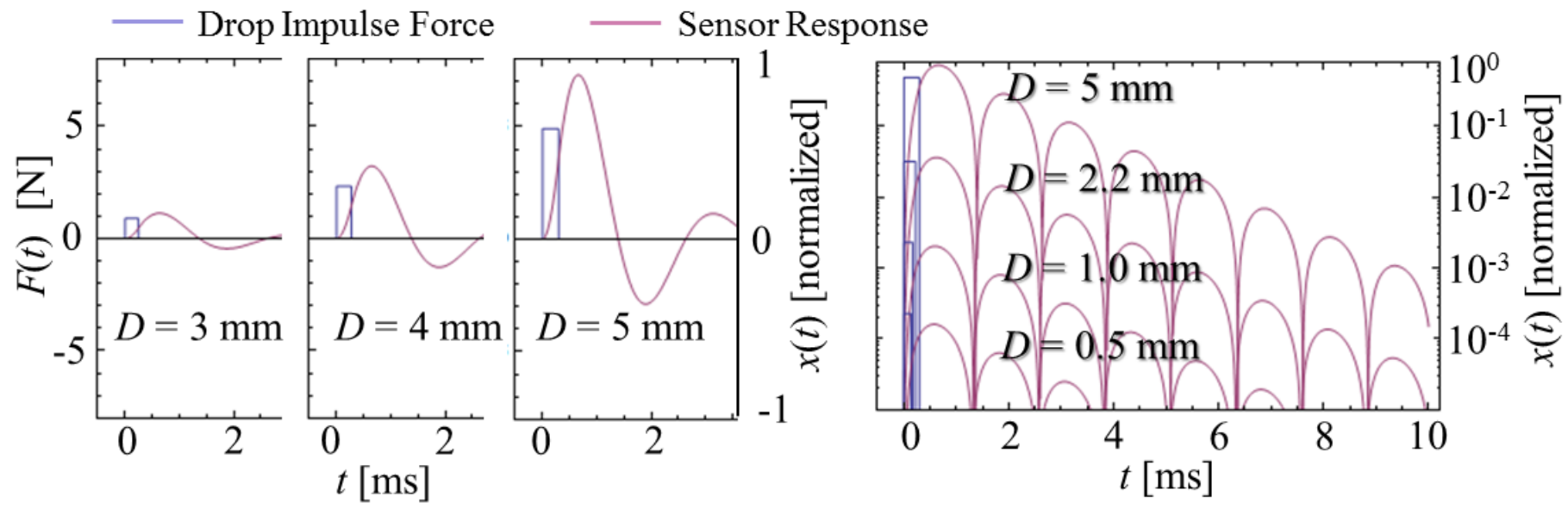

Fig. 3. SDOF model Equation (3) with, $f_0$ = 420 s$^{-1}$, $\rho$ = 1000 kg m$^{-3}$, $\gamma$ = 1500 s$^{-1}$, *M* = 0.01 kg, *e* = 1, and $\xi$ = 0.65.

## 5. DIGITAL SIGNAL PROCESSING SECTION

The signal processing section consists of multiple processing blocks, some of which are optional (see left side of Figure 4).

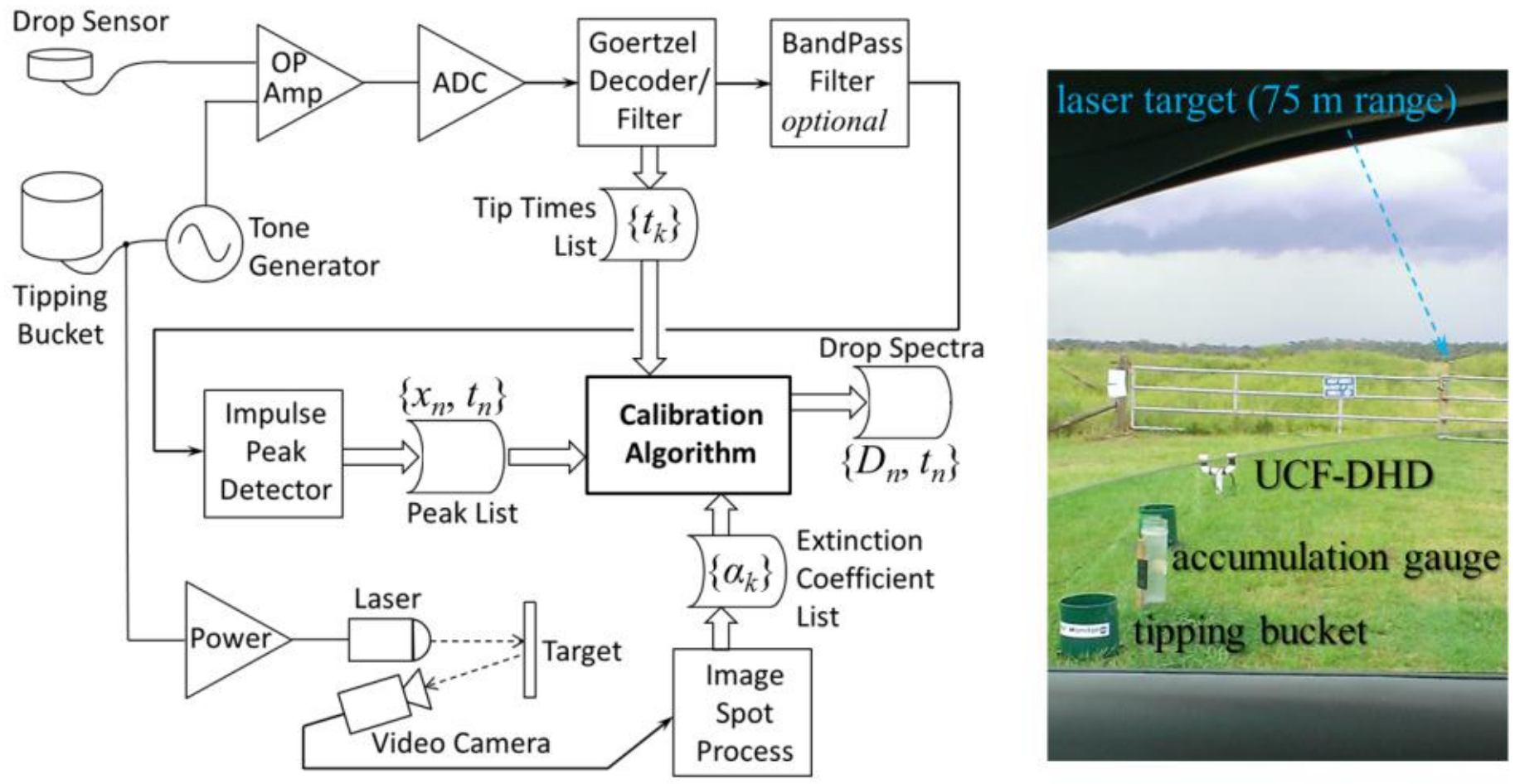

Fig. 4. Left: block diagram of signal processing electronics; Right: laser and camera with all electronics inside of the research vehicle, powered by 12V battery, with tipping buckets, rain gauge, and UCF-DHD shown through partially opened window.

### 5.1 Goertzel Algorithm

As shown in the left side of Figure 4, the top left corner is the disdrometer sensor. The analog signal is mixed with a tone pulse triggered by the tipping bucket (TB). The tone pulse width is very short compared to the time between tips and therefore does not degrade the drop spectrum measurement. The Goertzel decoder separates the tip tones from the sensor signal and creates a list of tip times. The tip times list, $\{t_k\}$, yields rainfall rate. The band pass filter section consists of optional filtering stages implemented as user selectable $N$th order low pass and high pass filters. Different processing strategies determine the cut-off frequencies of these filters relative to the resonant frequency of the sensor. The resonant frequency of the sensor is mostly determined by the encapsulant properties, primarily hardness.

### 5.2 Peak Detector

The peak detector creates a list of impulse maximums versus time, $\{x_n, t_n\}$, for all drops measured, where $x_n$ (not to be confused with $x(t)$) is proportional to the maximum amplitude of impulse $s(t)$. The number of impulses per second can easily range from 1 to 30 depending on rainfall rate and type of rainfall for the sensors tested. The total number of drops, or drop flux, is proportional to the area of the sensor. The drop impulse width, as shown in Figure 3, is dependent on the sensor characteristics, and again is mostly determined by the encapsulant material properties. A typical impulse width is dependent on the drop size, and for the largest drops (5-6 mm), 30 ms might be required for the impulse to fall below a noise threshold. For extreme rainfall rates, the flux may exceed 30 drops per second and as one can see, the trade-off between sensor size and coincidence of drop impulses sets the size of the sensor area to something in the 50 $cm^2$ range. The biggest challenge of the peak detector is to detect all impulses, while at the same time, avoid counting false impulses from the tail of a large drop (splashing from large drops can also lead to false counts).

### 5.3 Laser Spot Processing

The laser/camera system is triggered by the TB for convenience. This is not a requirement, but results in a simpler processing methodology. Each video camera image corresponding to tip time $k$, is converted to a spot region with an average greyscale value. For the 5mW 532 nm laser used in this work, the green component of the image is most sensitive to the laser, whereas the red and blue components are good indicators of background noise. The following image processing algorithm is applied to each $k$th frame, pixel by pixel:

$$F_{mn} = \left|\left(G_{mn} - R_{mn}\right)\left(G_{mn} - B_{mn}\right)\right|^{1/2}, \tag{6}$$

where, $G_{mn}$ is the 8-bit green value at pixel location (*m*, *n*), $B_{mn}$ is the blue value, and $R_{mn}$ is the red value. This algorithm converts the RGB color to a greyscale intensity $F_{mn}$. The filtered value is averaged over the spot within a half maximum intensity diameter (Lane *et al*. 2013). The diameter of the spot is also recorded, but only the intensity data is used in the final calibration. The intensity is then converted to extinction coefficient $\alpha_k = \ln(I_k / I_0)/2L$ where $2L$ = 150 m is the round trip distance from laser to target, $I_k$ is the average value of $F_{mn}$ over the laser spot, and $I_0$ is the value for no rain (see Figure 5). The video sequence from the camera imaging the laser spot is processed by this *image spot processing* algorithm which is based on the green filter of Equation (6). Figure 4 shows the output of the laser/camera system as a list of extinction coefficients $\alpha_k$ in units of [m$^{-1}$].

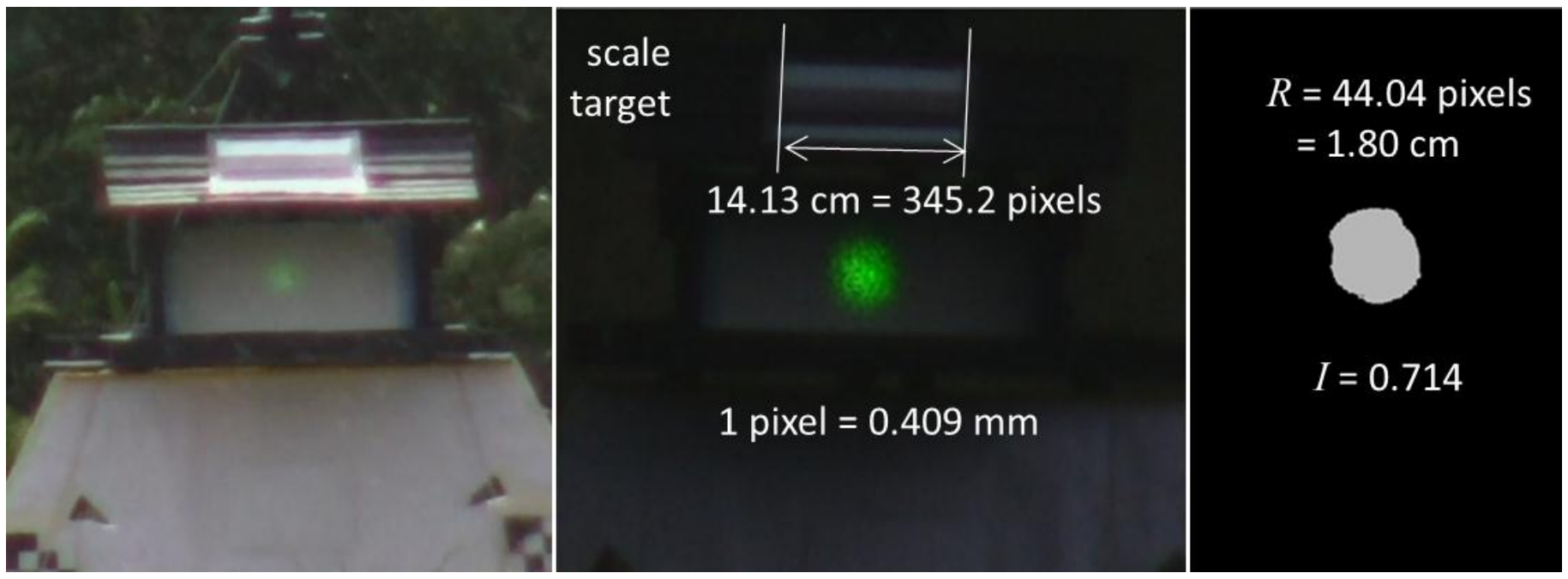

Fig. 5. Left: laser spot before rain; Middle: spot during rain; Right: output of image processing. Distance from laser/camera to target, $L$ = 75 m.

## 6. DISDROMETER CALIBRATION ERROR SURFACE

For any instrument that measures a physical quantity, verification and/or calibration is often based on comparison to data reported by a different instrument measuring the same quantity. For example, one or more collocated tipping bucket rain gauges are routinely used to verify calibration of a disdrometer. Similarly, comparison to other collocated disdrometers would provide a means to determine the quality of performance of the *drop distribution meter* (disdrometer) under test. The process of routinely verifying a disdrometer's calibration can be compared to the process of calibrating a disdrometer for the first time.

A central premise in this work is based on the understanding that a disdrometer's calibration is routinely verified by comparison of its derived *n*th moment to a collocated instrument that measures the same moment. Therefore it may be reasonable to calibrate a disdrometer by the reverse process, avoiding a factory single drop calibration procedure altogether. To demonstrate this concept, it is useful to consider an ideal simulation experiment using an ideal DSD, the exponential distribution, $N(D) = N_0 \exp(-D/D_0)$. The results are similar if the more general *gamma distribution* is substituted.

### 6.1 Two-Parameter Error Function

The first step is to define an error function, characterized by an error surface in multi-dimensional parameter space. For the tipping bucker/laser extinction case, the calibration error function can be defined as (where $\gamma$ and $\lambda$ are new variables not associated with Section 4):

$$E(N_0, D_0, a_A, \gamma, \lambda) = \lambda \sum_{k=1}^{M} \left(1 - \frac{\hat{R}_k(a,\gamma)}{R_k(N_0, D_0)}\right)^2 + (1-\lambda) \sum_{k=1}^{M} \left(1 - \frac{\hat{\alpha}_k(a,\gamma)}{\alpha_k(N_0, D_0)}\right)^2, \tag{7}$$

where $\lambda$ is a weighting factor (a real number between 0 and 1), $a$ and $\gamma$ are disdrometer calibration coefficients, and $k$ corresponds to the $k$th rain bucket tip. $\hat{R}_k(a,\gamma)$ is the disdrometer derived rainfall rate at the $k$th tip time and $R_k(N_0, D_0)$ is the actual (measured or simulated) rainfall rate. Likewise, $\hat{\alpha}_k(a,\gamma)$ is the disdrometer derived optical extinction coefficient, where $\alpha_k(N_0, D_0)$ is the measured opti-cal extinction using a laser/camera system (or any other transmissometer system). It is assumed (for this simulation experiment) that the disdrometer response can be characterized completely by a power-law calibration model:

$$\hat{v}_{ik} = a\, x_{ik}^{\gamma} \quad [\mathrm{m}^3] \quad , \tag{8}$$

where $x$ is an $N$-bit digital value represented by a fractional number between 0 and $1\text{-}2^{-N}$, which is the raw disdrometer output due to the impact of a drop of diameter $D_{ik}$, with an equivalent spherical volume, $v_{ik} = \frac{\pi}{6} D_{ik}^3$. The raw measured drop value $x_{ik}$ may represent the maximum amplitude of the impulse, the absolute value of the area under the impulse curve, or something else, depending on the disdrometer's processing details. The subscripts are used to account for the $i$th drop impulse occurring during the $k$th tip time. The dynamic range of the disdrometer is theoretically $2^N$, but in practice, digital systems are more often represented by a lower dynamic range, such as $2^{N-2}$. For the 16-bit system described in this paper, a dynamic range of $10^4$ is achievable.

For convenience, a power law form of drop terminal velocity will be assumed: $v_D(D) = \mu\, D^{\chi}$. Now the terms in Equation (7) can be evaluated by integrating the appropriate quantities over the DSD:

$$\begin{aligned} R_k(N_0, D_0) &= \int_0^{\infty} \tfrac{\pi}{6} D^3\, v_D(D)\, N(D)\, dD \\ &= \tfrac{\pi}{6} \mu\, N_0\, \Gamma(4+\chi)\, D_0^{4+\chi} \end{aligned} \quad , \tag{9}$$

$$\begin{aligned} \hat{R}(a,\gamma) &= \int_0^{\infty} a\, x^{\gamma} v_D'(x) N'(x)\, dx \\ &= (\pi/6)^{\gamma/\gamma_0}\, a\, a_0^{-\gamma/\gamma_0} \mu\, N_0\, \Gamma(1+3\gamma/\gamma_0+\chi) D_0^{1+3\gamma/\gamma_0+\chi} \end{aligned} \quad . \tag{10}$$

The parameters $a_0$ and $\gamma_0$ are a simulated drop to impulse transformation based on inverting Equation (8). The goal is to locate $a$ and $\gamma$ by examining the error surface associated with Equation (7). Thus when $a \to a_0$ and $\gamma \to \gamma_0$, a successful (simulated) calibration can be declared. Note that the primed variables denote a transformation from $D$ to $x$:

$$v_D'(x) = v_D(D)\big|_{D \to D(x)} \quad , \tag{11}$$

where $D(x) = (6/\pi\, a_0)^{1/3} x^{\gamma_0/3}$. The transformed DSD becomes:

$$N'(x) = N(D)\big|_{D \to D(x)} \frac{dD}{dx} \quad . \tag{12}$$

The remaining quantities in Equation (7) are computed as follows (using $Q_e = 2$):

$$\begin{aligned} \alpha_k(N_0, D_0) &= \tfrac{\pi}{4} Q_e \int_0^{\infty} D^2\, N(D)\, dD \\ &= \pi\, N_0 D_0^3 \end{aligned} \quad , \tag{13}$$

$$\begin{aligned} \hat{\alpha}_k(a,\gamma) &= \tfrac{\pi}{4} Q_e \int_0^{\infty} \left(\tfrac{6}{\pi} a\, x^{\gamma}\right)^{2/3} N'(x) dx \\ &= 2^{-\frac{2\gamma+\gamma_0}{3\gamma_0}}\, 3^{-\frac{2\gamma-2\gamma_0}{3\gamma_0}}\, \pi^{\frac{2\gamma+\gamma_0}{3\gamma_0}}\, a_0^{-\frac{2\gamma}{3\gamma_0}}\, a^{\frac{2}{3}}\, N_0\, \Gamma\!\left(1+\frac{2\gamma}{\gamma_0}\right) D_0^{\frac{2\gamma+\gamma_0}{\gamma_0}} \end{aligned} \tag{14}$$

As a check, when $a \to a_0$ and $\gamma \to \gamma_0$ , Equation (10) reduces to Equation (9) and Equation (14) reduces to Equation (13), as they should.

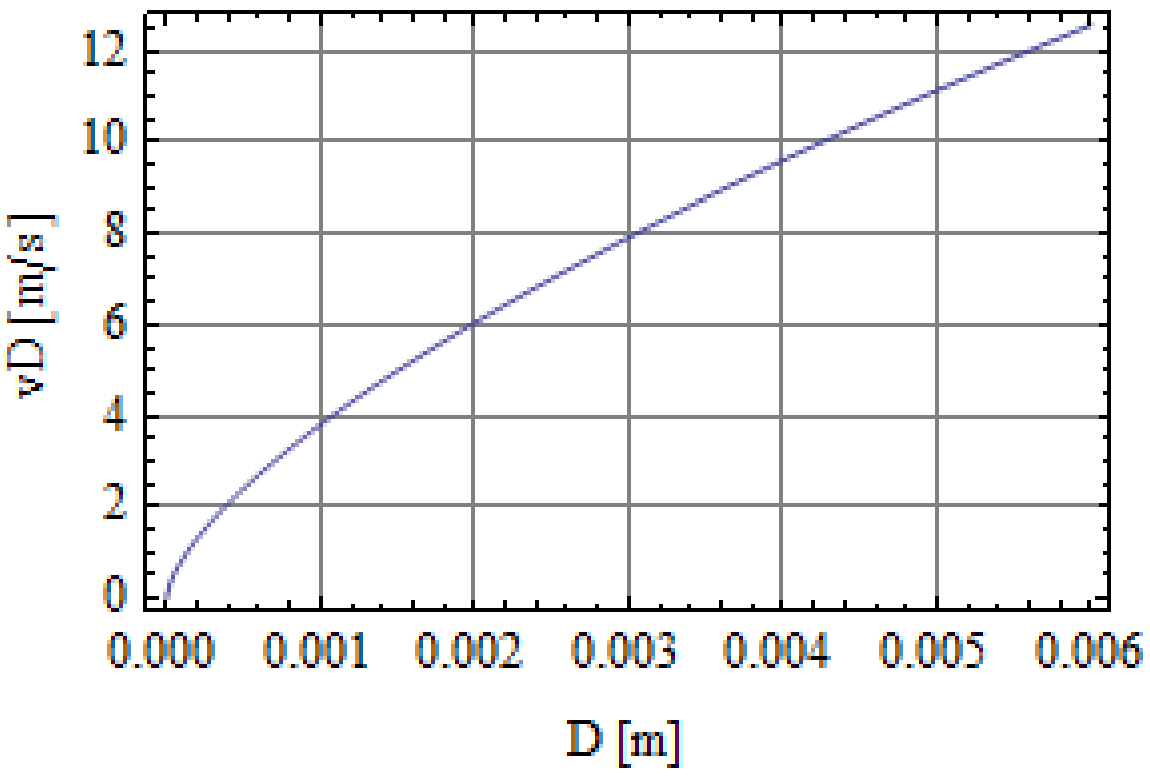

Fig. 6. Drop terminal velocity, as given by the power law $v_D(D) = 381\, D^{2/3}$ [m s$^{-1}$], with $D$ expressed in [m].

It is helpful to plot terminal velocity $v_D(D)$ , as shown in Figure 6. The simulated disdrometer response is plotted in Figure 7, using Equation (8) with $a_0$ = $10^{-6}$ [m$^3$] and $\gamma_0$ = 1.1, representing a realistic disdrometer output impulse to drop size transformation. Note that the $10^4$ dynamic range shown in the plot corresponds to a drop size range of 0.3 mm to 8 mm.

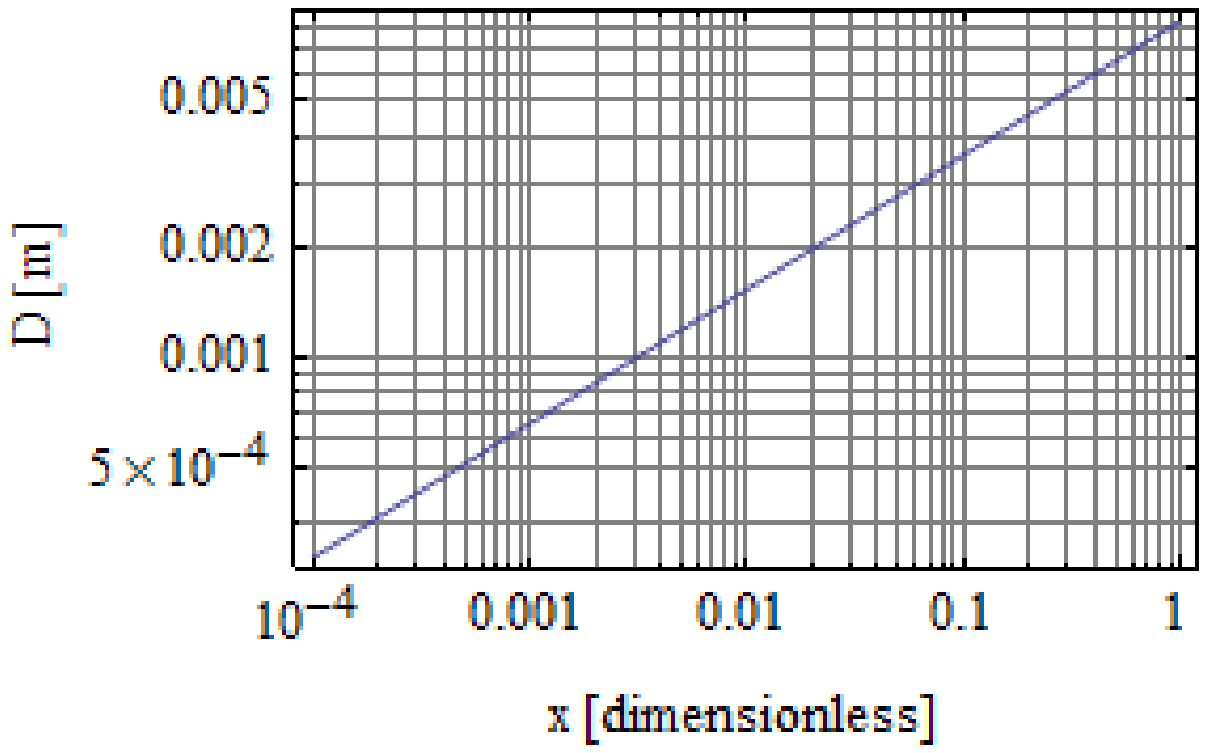

Fig. 7. Simulated disdrometer transformation curve.

The disdrometer calibration error surface can now be examined in detail using Equation (7). Figure 8 shows a case corresponding to a Marshall-Palmer (MP) like exponential DSD with $N_0$ = 8×10$^6$ [m$^{-3}$ m$^{-1}$] (Marshall *et. al*. 1948) and a rainfall rate from Equation (9) of $R$ = 130 [mm h$^{-1}$]. Three cases are shown. The left plot is the calibration error surface due to rainfall rate (tipping bucket rain gauge) only where $\lambda$ = 1. The middle plot is created with $\lambda$ = 0 and corresponds to the second moment measurement (laser extinction) only. The case on the right includes both the 11/3 moment (rainfall rate) and second moment (optical extinction) using $\lambda$ = ½ .

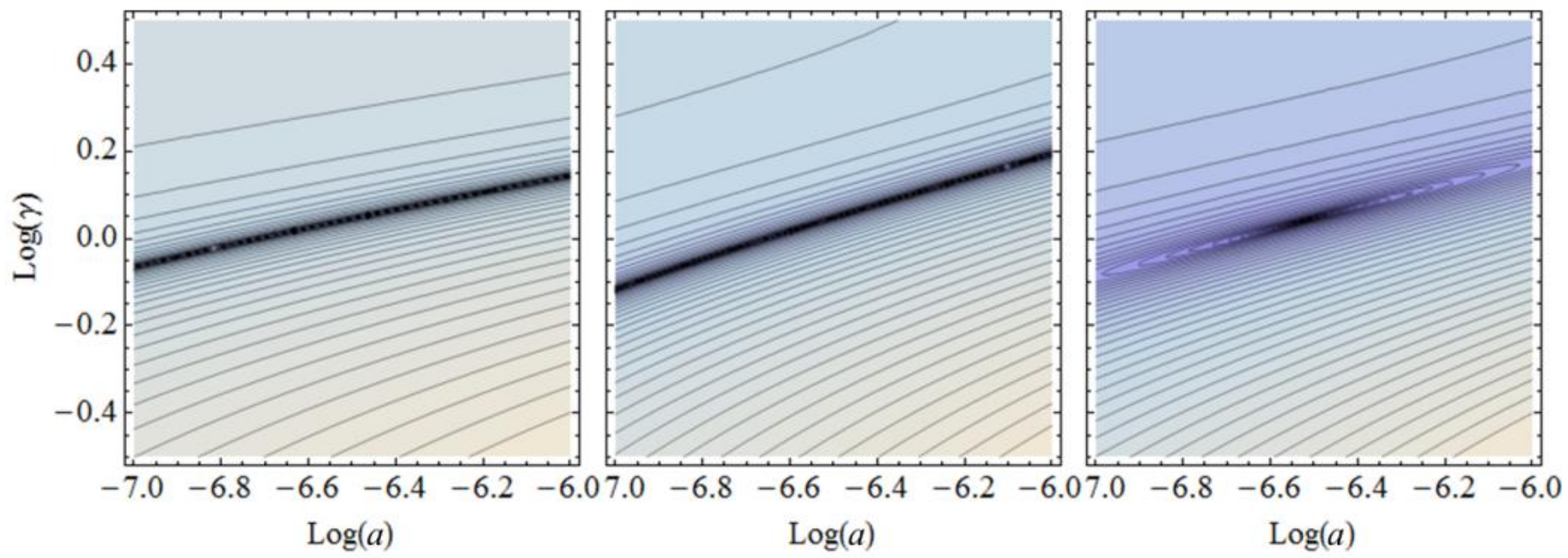

Fig. 8. Equation (7) with $N_0$ = 8×10$^6$ [m$^{-3}$ m$^{-1}$] and $D_0$ = 6.7×10$^{-4}$ [m], corresponding to $R$ = 130 [mm h$^{-1}$]. (left) $\lambda$ = 1; (middle) $\lambda$ = 0; (right) $\lambda$ = ½.

A second case shown in Figure 9 corresponds to a DSD with significantly fewer small drops than the MP DSD. This type of DSD may be associated with *impulsive rainfall* (IR). Impulsive rainfall can be defined as rapidly occurring and relatively short-lived precipitation events, associated with isolated convective thunderstorms common in Florida during the mid-summer months. Though not a requirement, an IR DSD is often characterized by a drop spectra flatter than the typical MP DSD, with a $D_0$ much larger than typical (Lane *et. al*. 2000). In this paper, an IR DSD is defined as a drop spectra with a flatter than normal size dependence. This shape may be the consequence a high degree of gravitational sorting, where smaller drops are stripped from the DSD aloft due to advection effects, high evaporation, significant updrafts, or a combination of these effects.

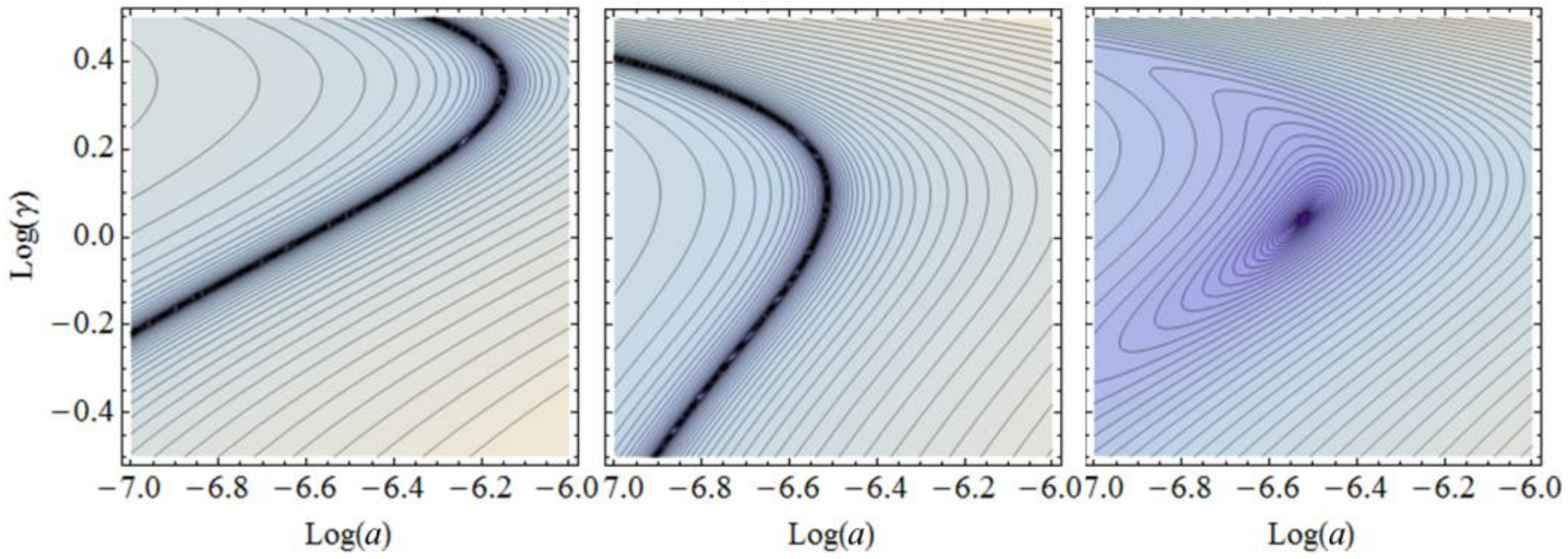

Fig. 9. Equation (7) with $N_0$ = 5×10$^4$ [m$^{-3}$ m$^{-1}$] and $D_0$ = 1.8×10$^{-3}$ [m], corresponding to $R$ = 82 [mm h$^{-1}$]. (left) $\lambda$ = 1; (middle) $\lambda$ = 0; (right) $\lambda$ = ½.

The third case shown in Figure 10 is a combination of the previous MP DSD and IR DSD, each sequentially on for a simulated time interval of 60 s.

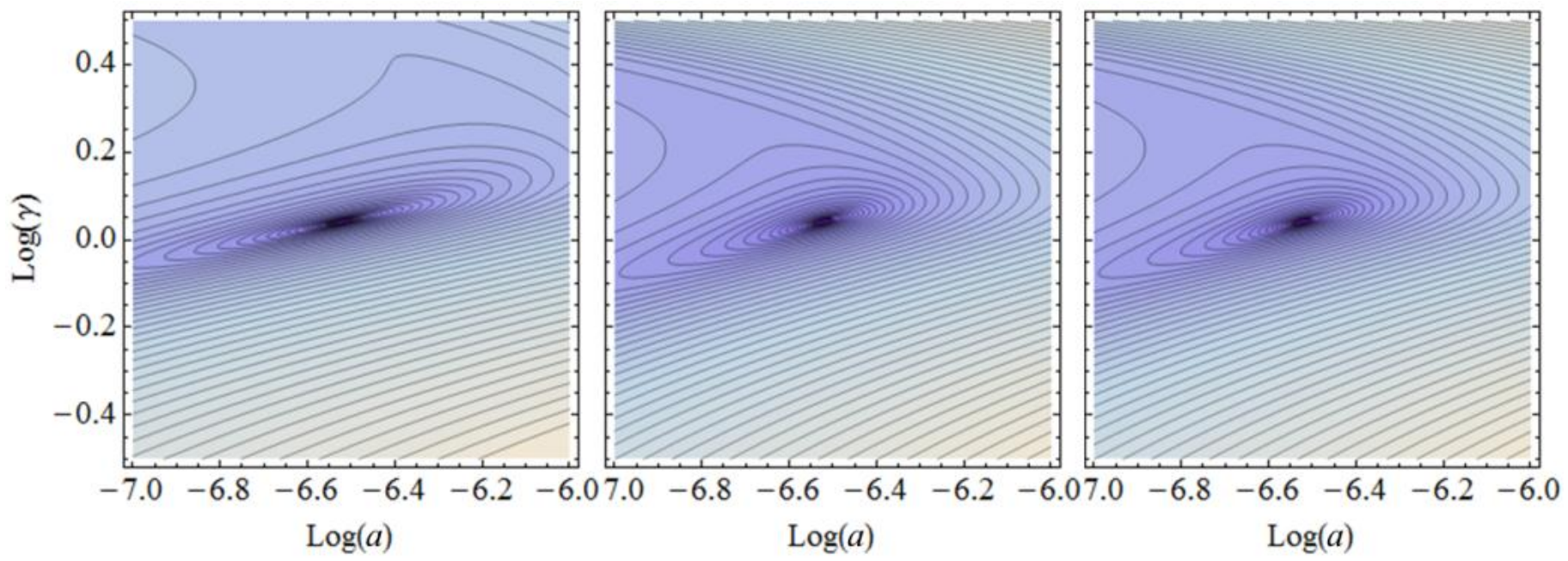

Fig. 10. Equation (7) with a mixture of 60 s at $N_0$ = 5×10$^4$ [m$^{-3}$ m$^{-1}$] and $D_0$ = 2.0×10$^{-3}$ [m], corresponding to $R$ = 137 [mm h$^{-1}$], followed by 60s at $N_0$ = 8×10$^6$ [m$^{-3}$ m$^{-1}$] and $D_0$ = 4.7×10$^{-3}$ [m], corresponding to $R$ = 25 [mm h$^{-1}$]. (left) $\lambda$ = 1; (middle) $\lambda$ = 0; (right) $\lambda$ = ½.

In the first case of Figure 8, it can be seen that disdrometer calibration during MP like rainfall using the 11/3 moment alone (rainfall rate) is not possible since the error surface has no well-defined minimum, only a valley minimum indicative of an infinite number of calibration solutions. The same is true of the optical extinction only error surface. The sum of the two error contributions also yields a trench like minimum, but with a defined minimum point. In this case the ability to find the true minimum is not ideal and is dependent on the "noise" in the measurement. Two typical sources of noise in the disdrometer calibration measurement are:

(1) Drops counted incorrectly - small drops can be missed if they occur directly after a large drop. A large drop may splash causing numerous erroneous small drop counts.

(2) Electronic and/or acoustic noise – this could be due to thunder, wind, or bad filtering on a power supply.

The second case shown in Figure 9 yields a more useable error surface for calibration, since location of the minimum (only in the $\lambda = ½$ case) is straight forward, even in the presence of noise. Unfortunately, the IR DSD associated with this case may only be found at particular locations and times of year. The third case of Figure 10 is a combination of the MP and IR DSD, which is more representative of a real impulsive rain event. In this ideal simulation case, a total of 0.11 [in] of rain is accumulated by the 120 s simulated rainfall event. Rainfall rate only, extinction only, or combination can be used to calibrate the disdrometer since the error minimum is well defined for all values of $\lambda$, but noticeably better in the extinction case where $\lambda < 1$.

Rainfall described by the MP DSD is more common than that described by the IR DSD, where a useable in situ calibration strategy may be devised by discriminating and using only the appropriate error surfaces, such as that shown in Figure 10, however it is desirable and practical to adaptively calibrate during all rainfall types.

### 6.2 Modified Error Function

In order to demonstrate an approach to this end, a modified error surface from Equation (7) is used:

$$E(N_0, D_0, a_A, a_B, \gamma) = \frac{1}{2}\sum_{k=1}^{M}\left(1-\frac{\hat{R}_k(a_A,\gamma)}{R_k(N_0,D_0)}\right)^2 + \frac{1}{2}\sum_{k=1}^{M}\left(1-\frac{\hat{\alpha}_k(a_B,\gamma)}{\alpha_k(N_0,D_0)}\right)^2, \tag{15}$$

where a third calibration parameter has been introduced by defining independent drop calibration model coefficients from Equation (8), $a_A$ for the rainfall rate term and $a_B$ for the optical extinction term. The solution methodology is not to locate the error surface minimum in three parameter space, but to vary the exponent parameter $\gamma$ in two parameter space until $a_B \to a_A$.

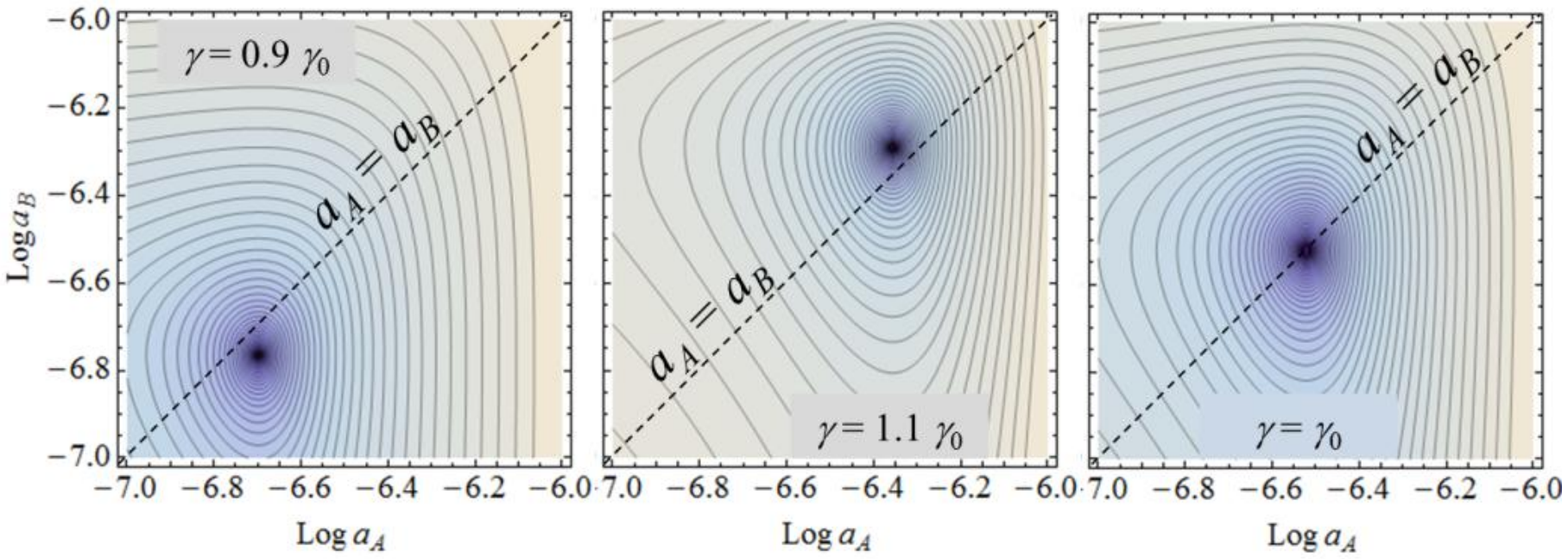

Fig. 11. DSD simulation example from Figure 8, using a modified error function given by Equation (15). The right plot represents the final solution after multiple iterations of the type shown in the right and middle plots.

A significant advantage of this approach is that an ill-formed error surface minimum, such as that shown in the right side of Figure 8, is transformed into a well-defined minimum as shown in Figure 11. A disadvantage of this strategy is that multiple error surfaces are computed during the iterative process of finding the final solution where $a_B \rightarrow a_A$. However, the increase in computational burden is a reasonable trade for the more significant benefit of utilizing ill-formed two-parameter error surfaces associated with most (and hopefully all) rainfall types.

## 7. EXAMPLE DATA PROCESSING

On August 3, 2013, between 21:00 and 22:30 UTC, data was collected at a site 17 km, 323.7° from the KLMB radar. Tipping buckets, accumulation gauge, UCF-DHD, and laser target were deployed outside of a vehicle (shown in right side of Figure 4). All electronics, including green laser, video camera, processing electronics, and audio recorder (for disdrometer), were inside of the vehicle and powered by a 12 V battery. The laser and camera were positioned so that the partially opened window does not interfere with the laser light beam. The vehicle and target were aligned to the approach of the oncoming storm so that the wind is generally opposite the partially opened window, thus minimizing the problem of rain damaging the electronics, laser, and camera.

### 7.1 Extinction Coefficient

The laser is turned on and off with a 50% duty cycle, 1 Hz square wave. The video camera records at 30 fps. When the tipping bucket tips, the laser is held on for 3 s, and an audio tone is mixed with the disdrometer audio channel. During image processing of the video stream, the first step is to decimate the sample rate to 10 fps. A section near the center of the image is cropped as the region of interest (ROI). Each image ROI is processed by Equation (6), pixel by pixel. This is essentially a green band pass filter which transforms the black and white laser target into a totally black image under normal solar illumination until the green laser spot appears. Figure 12 shows the output of the image processing filter Equation (6), but only for the 3 s tip regions (the 1 Hz pulses have been removed). During heavy rainfall, the output of Equation (6) is greater than zero (not totally black) when the laser is in the off portion of the cycle. This portion of the signal is captured and is treated as a background back-scatter part to be removed from the high intensity part. The fact that the signal appears as a back-scatter signal at the output of the green band pass filter can be explained empirically by assuming that rain backscatter is shifting the solar spectrum to the green. This effect is clearly related to higher rainfall rate which creates higher backscatter before the laser reaches the target.

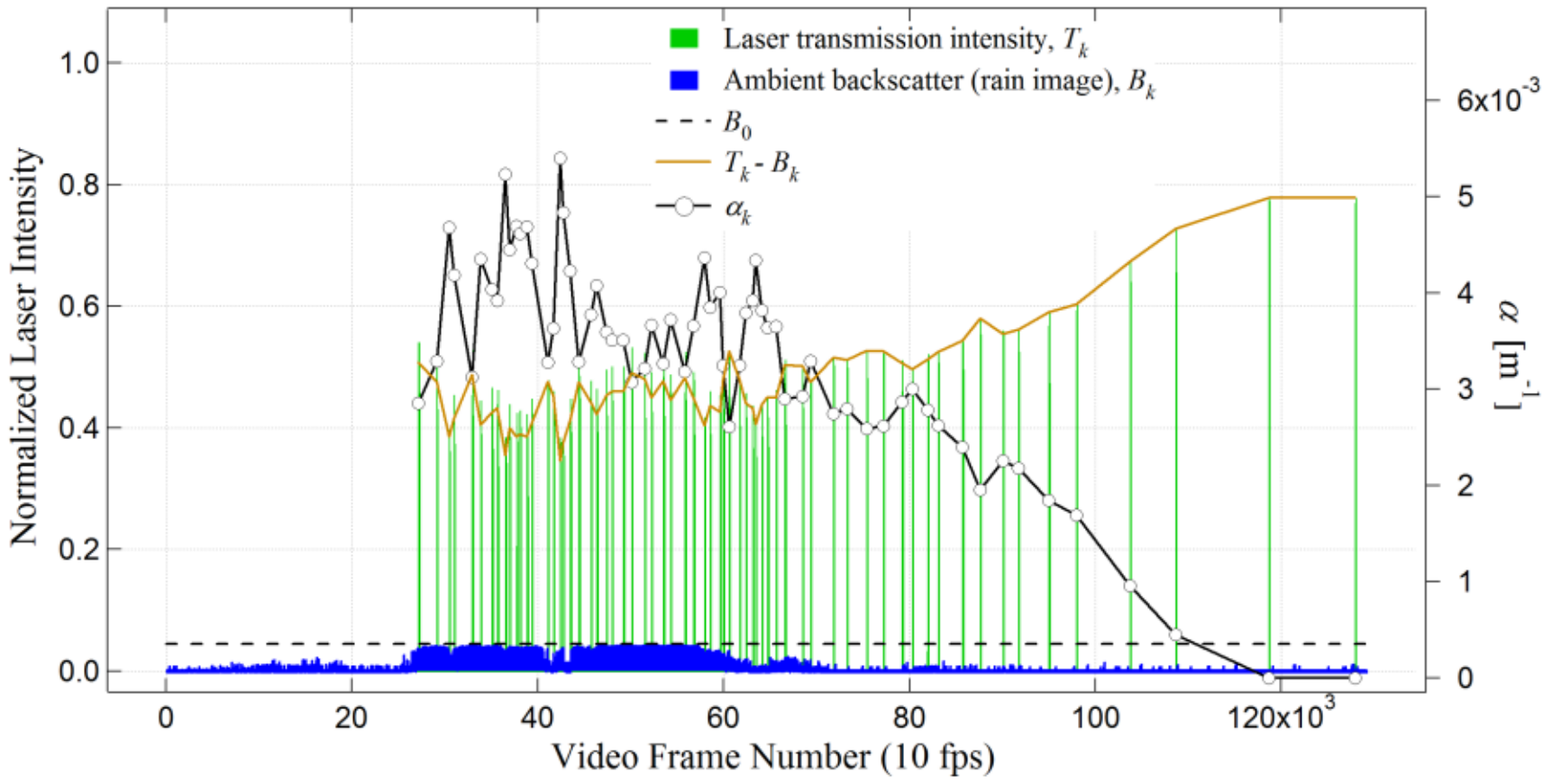

Fig. 12. Green lines represent the normalized laser intensity of the spot on the target viewed by the video camera, where the round trip distance $2L$ = 150 m, and each vertical line corresponds to a rain gauge bucket tip. The blue lines at the bottom are estimates of the back scatter from the rain, which increases with increased rainfall rate. This background is subtracted from the intensity (green lines), then converted to extinction coefficient, $\alpha$, shown by the solid black line where open circles, also corresponding to TB tip times.

Referring again to Figure 12, the green lines represent the normalized laser intensity viewed by the video camera on the target, processed by Equation (6), where the round trip distance is $2L$ = 150 m, and each vertical line corresponds to a rain gauge bucket tip. The blue region at the bottom of Figure 12 is an estimate of the back scatter from rain, which increases with increased rainfall rate. This background is subtracted from the transmitted intensity (green lines), then converted to extinction coefficient $\alpha$ as described in Section 5.3, shown by the black line with open circles.

### 7.2 Disdrometer Calibration

The modified error function from Equation (15) and the procedural methodology described in Section 6.2 leads to a disdrometer calibration algorithm (see Appendix A for additional details). The left side of Figure 13 shows the error surface based on Equation (7), where the minimum is not well defined. More will be said about this in the Summary Section, but what appears to be the primary indicator of a well-defined versus a poorly defined error surface minimum is the DSD flux distribution shape. A *broadband* drop spectra (spectrum which contains a more uniform mix of large and small drops) appears to generate a well-defined two parameter error surface from Equation (7). The August 3, 2013 rainfall event of this example appears to be characterized by down drafts with no sorting of drop sizes, leading do a more typical *narrowband* drop spectra (drop flux spectrum is peaked at one particular drop diameter, typical of most MP or gamma distributions).

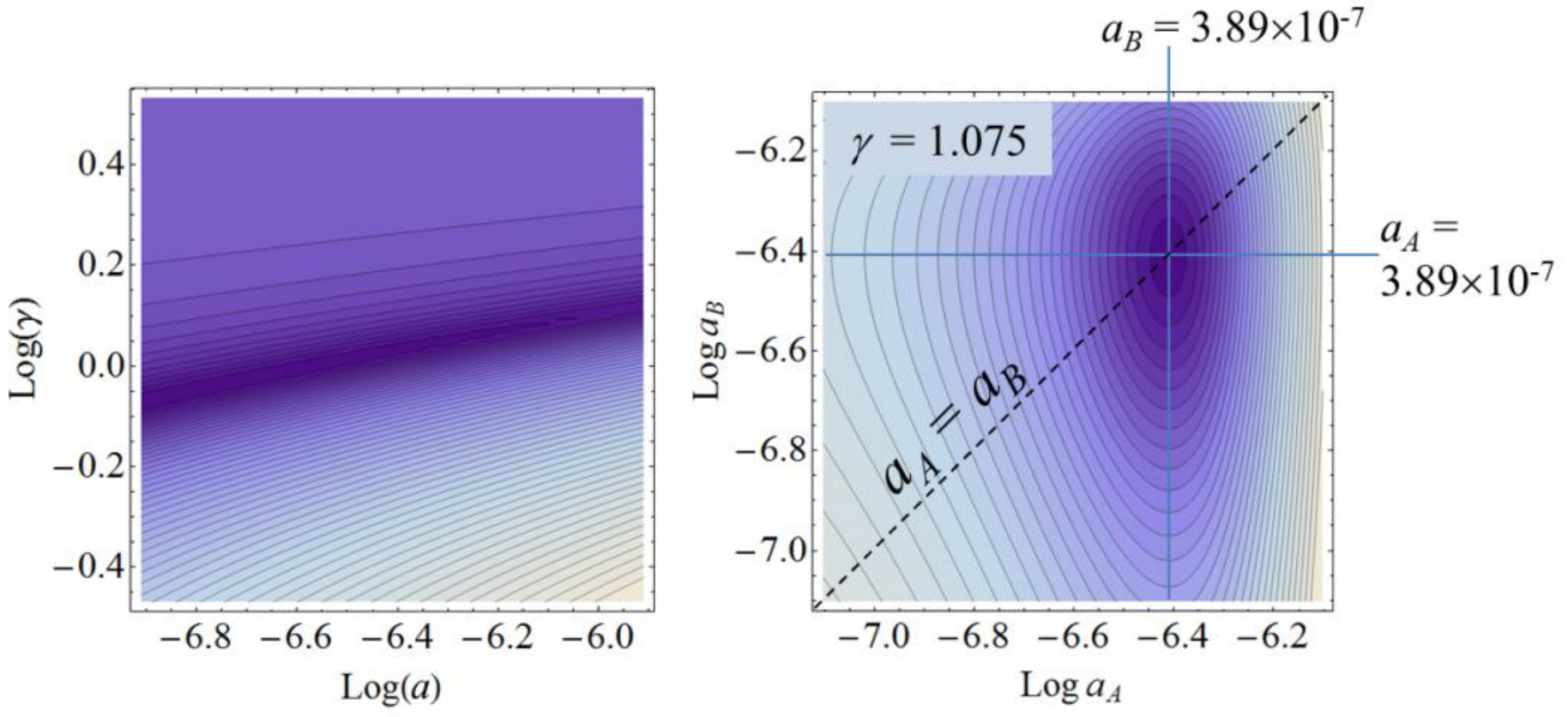

Fig. 13. Disdrometer calibration using dataset of August 3, 2013. (left) the error surface based on the concepts of Equation (7), where the minimum is not well defined; (right) modified error function based on the concepts of Equation (15) with a well-defined minimum.

The modified error function based on the conceptual approach of Equation (15), as shown in the right side of Figure 13, clearly shows a well-defined minimum and provides unambiguous calibration coefficients. The error function used to generate the plots in Figure 13 is based on Equation (15), but the actual error function used is a modified version necessary to process disdrometer drop spectra data, based on definitions and notation described in Appendix A:

$$E(a_A, a_B, \gamma) = \sum_{k=1}^{M}\left(1 - \frac{a_A X_k^{(1)}}{V_0}\right)^2 + \sum_{k=1}^{M}\left(1 - \frac{(6/\pi\, a_B)^{(2-\chi)/3} Y_k^{(1)}}{U_k}\right)^2 . \qquad (16)$$

Another way to illustrate the solution based on Figure 13 and Equation (16) is shown in the plot of Figure 14. This plot shows the valley minimum of the left side of Figure 13, with $a_A$ versus $\gamma$ and $a_B$ versus $\gamma$. The algorithm described in Appendix A is a general numerical method for solving this equation where the disdrometer model can be defined as series of polynomial terms in $x$. Even though the Appen-dix A method is greatly simplified with the single term model definition by Equation (8), it is useful and convenient (since it was previously coded as a Fortran function) to use the full matrix implementation of Appendix A. Using that approach, or any other equivalent method, Figures 14 shows multiple solutions for $a_A$ and $a_B$ for various values of $\gamma$. Since the goal is to find the $\gamma$ where $a_A \rightarrow a_B$, plotting $e = a_A - a_B \rightarrow 0$ clearly shows the desired value of $\gamma$. That point is equivalent to the error minimum in the right side plot of Figure 13. The method shown in Figure 14 is computationally more efficient since it does not involve creating a large number of error calculations for each iteration of $\gamma$. But Figure 13 conveys the concept more clearly since it is a direct plot of the error surface described by Equation (16). The calibration procedure described above results in $D = 9.06\, x^{0.36}$ [mm]. The quantities plotted in Figure 15 are generated from $D(t)$, where $D(t)$ is equivalent to all $D_{ik}$ (the calibrated version of $x_{ik}$) using standard calculations (Uijlenhoet *et al*. 2011, Atlas and Ulbrich 1977).

Appendix A discusses one (of many) possible methods for processing the error function to locate the minimum and extract the disdrometer calibration parameters. However, the only way to verify calibration is to compute the disdrometer derived *n*th moment values that can be compared with the equivalent collocated instruments that measure the same quantities. This is shown in Figure 15 where the thin lines represent the actual collocated measurement, and the thick lines represent the disdrometer (after calibration) computed equivalent *n*th moment values. The green lines (bottom line set) show the optical extinction comparison. The blues lines (middle line set) show the rainfall rate tipping bucket comparison. The upper line set (black lines) show a comparison of radar reflectivity values. Note that in this case the comparison reflectivity is not from KMLB radar, but is computed using an exponential DSD as described by Lane, *et al*. (2013). The Melbourne NWS KMLB radar showed a very low reflectivity, not consistent with a typical Z-R relation. For this reason, it was not plotted in Figure 15, and was one of the clues that this particular rainfall event was perhaps dominated by downdrafts (Ahammad, *et. al*. 2002) and characterized by a typical MP-like drop spectra, i.e., narrowband drop spectrum.

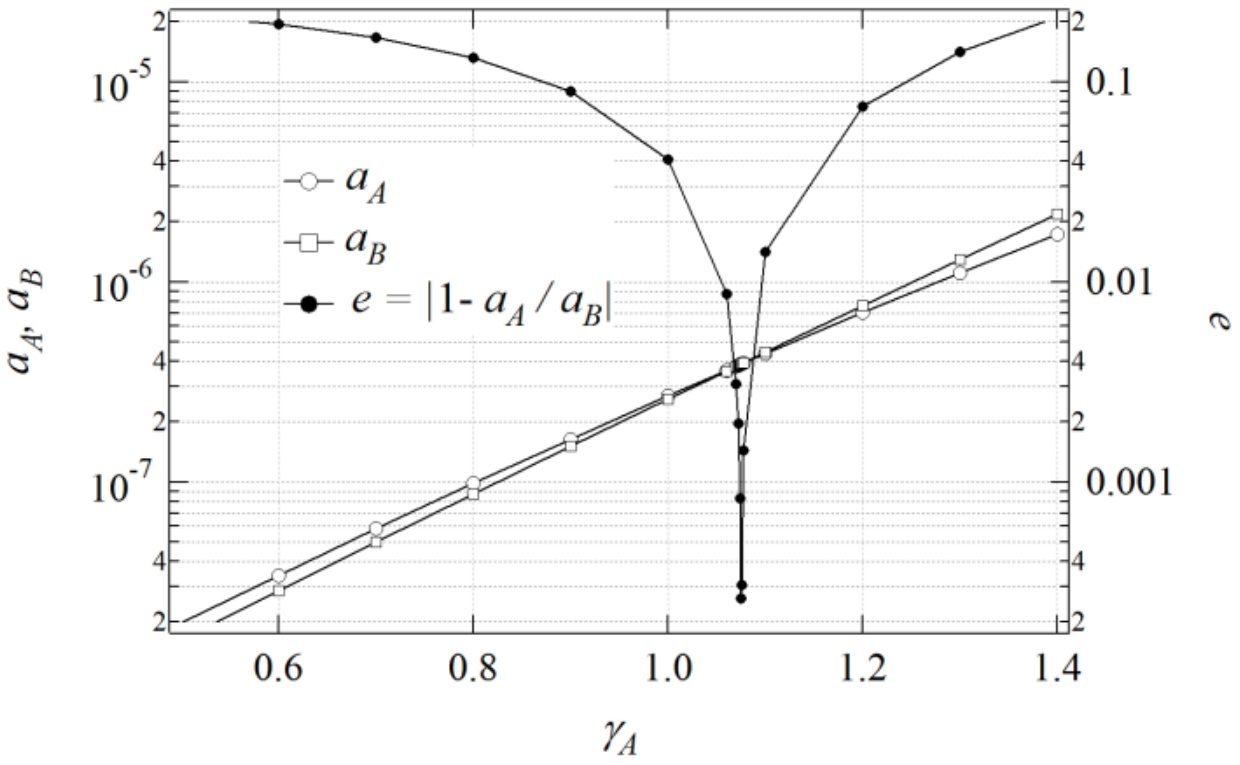

Fig. 14. This plot shows the valley minimum of the left side of Figure 13 with $a_A$ versus $\gamma$ and $a_B$ versus $\gamma$. The convergence point and solution is emphasized by plotting $e = a_A - a_B$ versus $\gamma$.

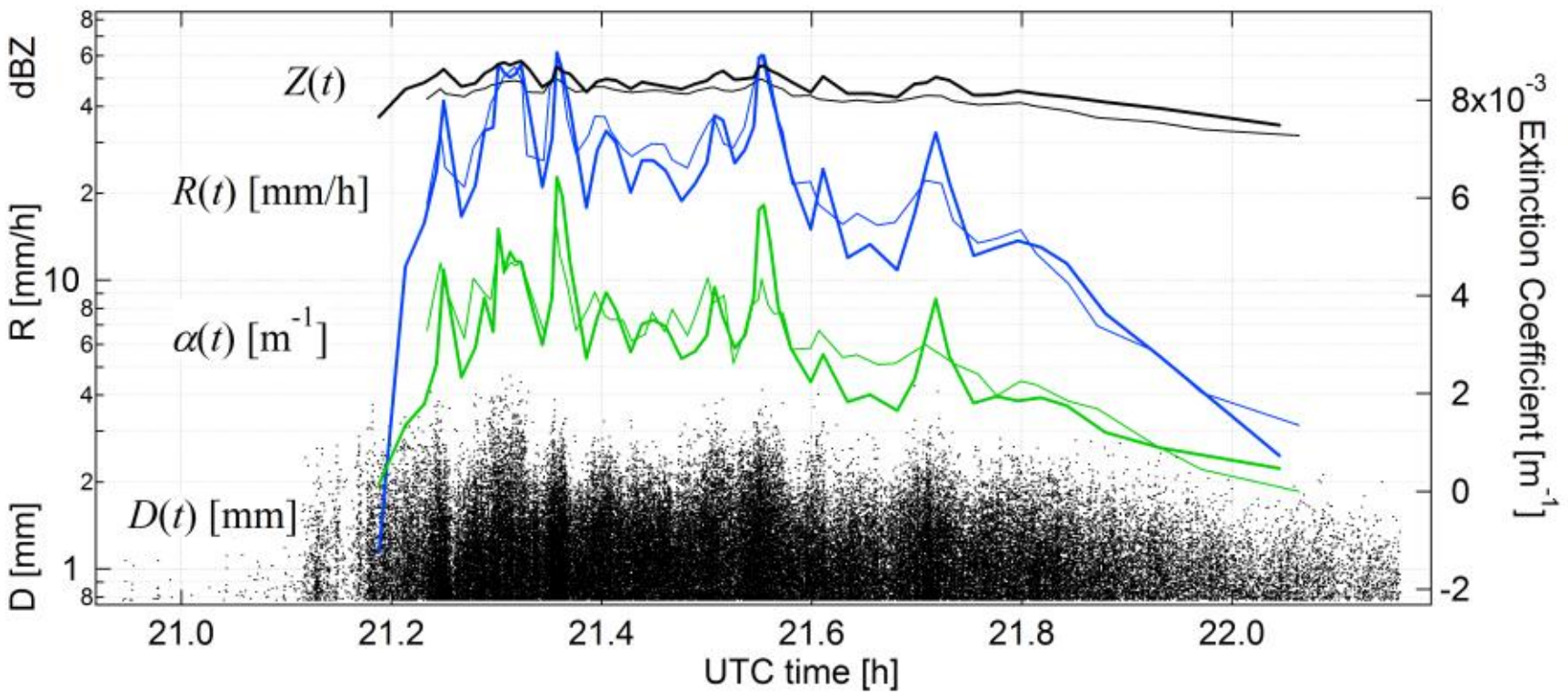

Fig. 15. DSD flux $D(t)$ plotted as black dots; green lines are extinction coefficient, thin line is from the laser measurement, thick line is derived from $D(t)$; blue lines are rainfall rate, thin line is from the TB measurement, thick line is derived from $D(t)$; black lines are computed radar reflectivity, thin line derived from the laser measurement and assumption of a pure exponential DSD model, thick line is derived from $D(t)$.

## 8. SUMMARY

Previous research suggests that in situ calibration may be a useful strategy towards implementation and deployment of low-cost disdrometers (Kasparis *et al*. 2010, Jong and Hut 2011). With this goal in mind, prototype disdrometers were developed, fabricated and tested, and are in operation at the University of Central Florida, Orlando, FL and the Cyprus University of Technology, Lemesos, Cyprus (see Figure 16). These disdrometers are constructed from off-the-shelf, low-cost parts and materials, and by eliminating the need for single drop calibration, the total system cost reduction may hopefully lead to the realization of dense disdrometer networks for the goal of studying spatial and temporal variability of hydrometeor size distributions.

In Section 4, Equations (2) through (5) represent an idealized SDOF model of a generic impact disdrometer. Equation (8) of Section 6 correspond to an empirical model constructed for the sole purpose of calibration. The SDOF model results in $x(D) \propto D^{3.67}$, whereas the calibration fit of Section 7 results in $x(D) \propto D^{2.8}$. This disdrometer response appears to be dependent on something slightly less than the mass of the drop, which is surprising since the usual expectation is that the an impact disdrometer response lies somewhere between momentum and drop kinetic energy dependence. The SDOF model predicts a response proportional to drop momentum.

The question of how often to calibrate can be addressed. On one extreme disdrometers could be calibrated at a factory facility during a few naturally occurring rain events, using tipping buckets and a laser combination. The calibration coefficients are then loaded and locked into disdrometer processing memory and the disdrometers are used without further adaptive calibration from that point on, wherever they are deployed. On the other extreme, tipping buckets and a laser can be used to continuously update calibration. A disdrometer is almost always collocated with one or more tipping bucket rain gauges, so that rainfall rate data is most likely available without extra cost and available for calibration on a continuous basis. The laser is more costly than a tipping bucket, primarily because of the camera and associated image processing. A laser generally has some inherent safety issues to consider, even though a class 3A laser was safely used in this work.

Continuous tipping bucket only calibration is possible and has been discussed previously (Kasparis *et al*. 2010). Simulations presented in Section 6 of this paper strongly suggest that extreme caution must be exercised in using a single moment calibration strategy. A laser only calibration is also possible (with similar precautions) but may have some advantages over a tipping bucket only calibration: quantization of the tip is a problem in low rain rates; tipping bucket sloshing is a problem in high rain rates. The second moment provides more detail for small drop sizes, while the larger drops ensure a better signal to noise performance, which should compensate for the shift of the second moment peak to the low end of the size distribution. However, an autonomous laser/camera/processing arrangement is certainly more costly than a tipping bucket calibration. Also, the tipping bucket calibration has one distinct advantage over the laser (in addition to lower cost) , rainfall rate and disdrometer data are both DSD flux measurements and do not require an approximation of drop terminal velocities for comparison.

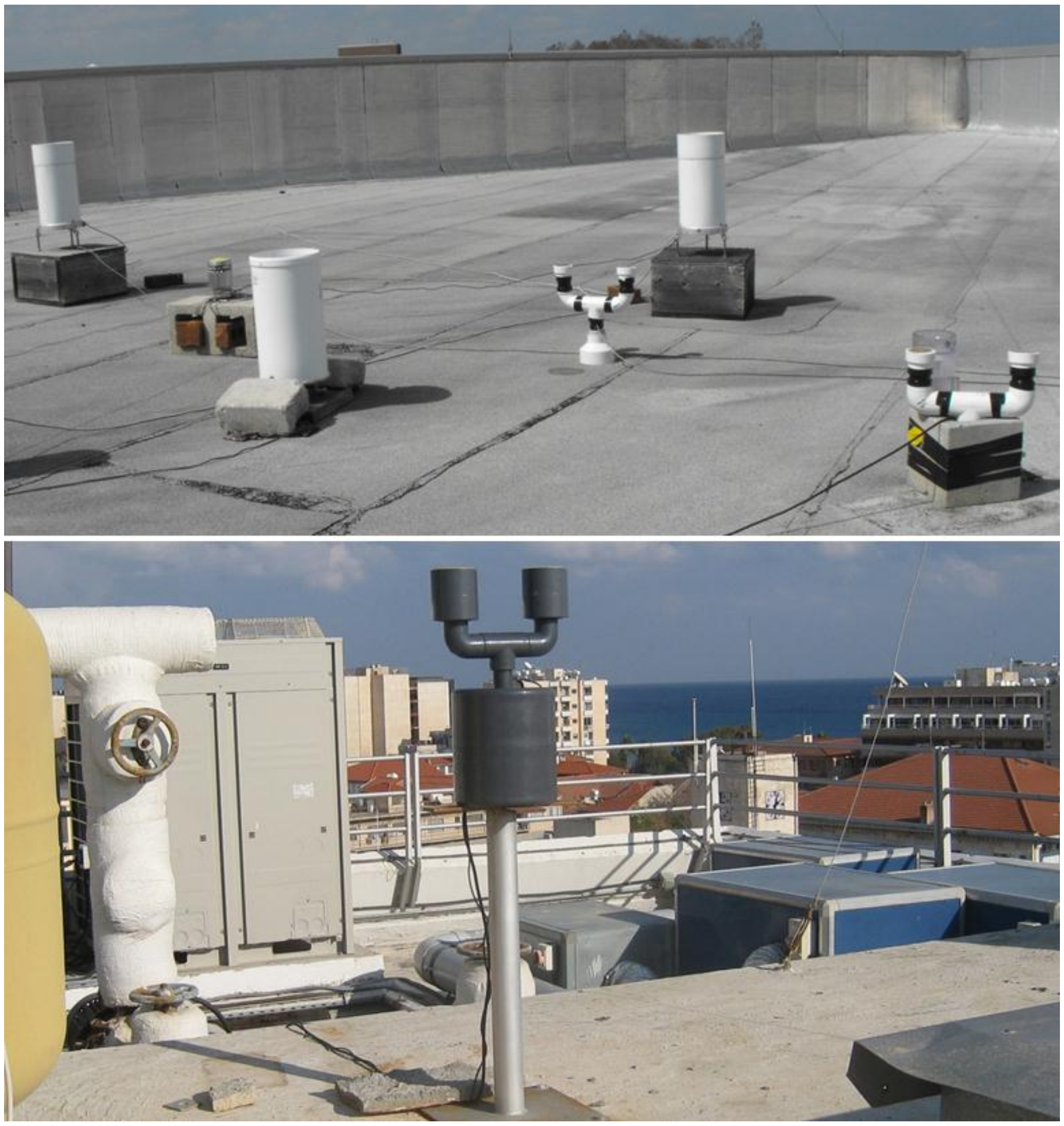

Fig. 16. Top: Roof of UCF Engineering Building, showing three tipping bucket rain gauges (white cylinders) and three disdrometers: Joss disdrometer on left, and two UCF-DHD disdrometers in center and far right. Bottom: CUT-DHD on the roof of the Cyprus University of Technology.

Problems with tipping bucket resolution on the low end or sloshing errors on the high end of rainfall rate are similar to the question of ideal laser extinction distance or ideal disdrometer sensing area. A tipping bucket opening can be made larger than then the standard 8 [in] diameter, making it more sensitive to lower rainfall rates. But then sloshing errors at high rainfall rates become more pronounced. Similarly, if the disdrometer sensing head is made larger, it will do better at capturing the drop spectrum for larger drops, but smaller drop measurements will suffer due to increased drop coincidence, a condition that cannot be processed correctly. Disdrometer saturation can be avoided by proper analog gain design, ensuring that the response curve accommodates all physical drop sizes, as shown by the example plot of Figure 7. This example will not saturate for drop sizes up to $D$ = 8 mm, well beyond those of physical rainfall drop diameters. However, it would saturate for all but the very smallest hail stones. An increased distance between the laser and target will increase the resolution of optical extinction measurements at lower rainfall rates, but higher rainfall rates may completely obscure the laser spot. Tradeoffs must be made based on laser power and wavelength, target distance, and camera sensitivity. For the consumer grade video camera and 5 mW green laser used in this work, a laser to target distance of $L$ = 75 m seemed to provide a reasonable compromise for resolution on the low end and sensitivity on the high end of rainfall rate.

The single moment calibration, i.e., tipping bucket only, optical extinction only, or radar reflectivity only, has been shown to be limited to only certain types of DSDs. Based on the simulation analysis of Section 6, it appears that the single moment calibration can only be successful during rainfall consisting of a larger than normal population of large drops. This type of DSD is characterized by a flatter drop spectrum than the typical MP or exponential DSD spectrum. Since the disdrometer is unaware of drop terminal velocity and actual spatial DSDs, and since it is measuring only DSD flux, this would suggest that single moment calibration could be successful during events consisting of updrafts and/or pronounced gravitational sorting. Investigating this relationship further may be an area of future work. Nonetheless, the use of two or more moments, such as the tipping bucket rainfall rate and laser optical extinction, seems to circumvent the need for ideal rainfall events for useful disdrometer calibration.

The methods described in this paper, with some modification, can also be applied to post processing disdrometer output from most commercial instruments, not limited to impact type disdrometers. The concept of in situ calibration is linked to the concepts of adaptive filtering (Widrow *et al*. 1985) where numerous algorithms have been developed to utilize a few basic concepts. One such approach common to adaptive filtering is to include weights in the error function summation terms to achieve specific goals. If a disdrometer for example was known to have saturation problems on the high end, thus underestimating the actual size of a large drop, weights could be included that were proportion to rainfall rate and/or optical extinction which would then bias the outcome to favor the calibration of larger drops.

No matter what strategy is prescribed, it is important to recognize that in situ disdrometer calibration strategies only guarantee that the final disdrometer derived $N(D)$ matches the calibration sources through the equivalent moments of the DSD, $\int D^n N(D)dD$. Disdrometer calibration and estimation of the drop size distribution aloft are based on the assumption that measuring the moments of the distribution is sufficient for estimating the distribution aloft (within a few meters above the disdrometer). The more moments that can be measured, the better the disdrometer calibration and estimation of $N(D)$ becomes.

Quality control of the calibration setup, a necessary requirement for valid disdrometer calibration, may include verification that the tipping buckets are correctly calibrated. This can generally be accomplished by comparing multiple tipping buckets collocated with multiple accumulation rain gauges. It is more difficult to validate the laser measurement. Independently checking camera response to a calibrated light source and checking the laser output with a calibrated photometer are standard techniques of validating a laser/camera system.

**ACKNOWLEDGEMENTS**: We gratefully acknowledge the ECE Division of the Electrical Engineering and Computer Science Department of the University of Central Florida, Orlando, Florida and the Cyprus University of Technology, Lemeso, Cyprus for use of their facilities. We also gratefully acknowledge the Cyprus Research Promotion Foundation for their support during the 2009-2011 phase of this project.

# APPENDIX A: Calibration Algorithm

The fundamental calibration goal is to locate the minimum in the error function parameter space as described in Section 6 and demonstrated in Section 7. This can be done graphically. However, in order to automate the process in an autonomous system, an efficient computational method is needed. The following calibration processing method is a direct extension of that described in Kasparis *et al.* (2010), but is just one of many possible approaches.

In the current method an additional laser extinction term is included in the *Kasparis* error minimization, as shown by Equation (7) and (15). Using the mathematical notation and computational approach of Metzger *et al.* (2010), the error function minimum can be easily found for a given $\gamma$. The *Metzger* notation provides a convenient and compact form, easing evaluation complexity, where a simple matrix inversion solves the problem directly. Of course we assume that the matrix inversion comes without a real-time computational price, which is not really true, but is a useful working assumption.

The $\mathbf{A}_k$ vector in Equation (A-1) corresponds to the sum of the tipping bucket framed (by the *k*th tip) disdrometer impulse amplitudes $x_{ik}$, as described in Section 6. The sum of the $X_k^{(j)}$ components over index *j* is proportional to the the sum of all $M_k$ single drop volumes that impact the disdrometer during the *k*th time interval. $V_0 = h_0 A_s$ is the total volume of water impacting the disdrometer during the *k*th tipping bucket time interval, and is equal to the tip depth ($h_0$ = 0.01 in) multiplied by the area of the disdrometer, $A_s$ = 58 cm$^2$. In the particular response model assumed in this paper, described by Equation (8), only the first component of $\mathbf{A}_k$ is non-zero. Therefore, $X_k^{(1)}$ is the sum of all drop volumes that impact the disdrometer during time interval *k*, and $\mathbf{A}_k$ essentially collapses to a scalar value, $A_k$. The ratio of disdrometer water volume to tipping bucket volume described by $A_k$ is proportional to the ratio of rainfall rates shown in the first term of the error function of Equations (7) and (15).

$$\mathbf{A}_k = \begin{pmatrix} X_k^{(1)}/V_0 \\ \vdots \\ X_k^{(N_A)}/V_0 \\ 0 \\ \vdots \\ 0 \end{pmatrix} , \quad \begin{matrix} X_k^{(1)} = \sum_{i=1}^{M_k} x_{ik}^{\gamma_A} \\ X_k^{(2)} = \sum_{i=1}^{M_k} x_{ik}^{2\gamma_A} \\ \vdots \\ X_k^{(N_A)} = \sum_{i=1}^{M_k} x_{ik}^{N_A\gamma_A} \end{matrix} \quad . \qquad \text{(A-1)}$$

Equation (A-2) describes the $\mathbf{B}_k$ vector due to the optical extinction data, also framed by the *k*th tip time interval. As in the previous case, $\mathbf{B}_k$ also essentially collapses to a scalar value $B_k$ for the case of the disdrometer response model specified by Equation (8). The quantity represented by $Y_k^{(1)}$ is proportional to the sum of the drop cross-sections, or in essence the second moment, corresponding to tip interval *k*. The exponent $\gamma_B = \gamma_A(2-\chi)/3$ transforms the drop volume related quantity in Equation (A-1) to a second moment related quantity in Equation (A-2), where $\chi$ = 2/3 is from the terminal velocity approximation used in Section 6 (Atlas *et. al*. 1977). The scaling factor, analogous to $V_0$ above, is $U_k = \mu A_S \Delta t_k M_{2_k}$ , where $\mu$ = 380.7 m s$^{-1}$ m$^{-2/3}$, (again from the terminal velocity approximation of Section 6), $\Delta t_k$ = time between *k* and *k*-1 TB tips, and $M_{2k}$ is the second moment of the DSD at tip *k*. $M_{2k}$ is calculated from the measured extinction coefficient, $\alpha_k = \pi Q_e M_{2_k}/4$, where $Q_e$ is the scattering efficiency factor for extinction (Berg *et al*. 2011). In this application $Q_e$ is assumed to be a constant equal to 2. The ratio of disdrometer drop second moment to measured optical extinction described by $B_k$ is proportional to the ratio of optical extinction coefficients shown in the second term of the error function of Equations (7) and (15).

$$\mathbf{B}_k = \begin{pmatrix} 0 \\ \vdots \\ 0 \\ Y_k^{(1)}/U_k \\ \vdots \\ Y_k^{(N_B)}/U_k \end{pmatrix} , \quad \begin{array}{l} Y_k^{(1)} = \sum_{i=1}^{M_k} x_{ik}^{\gamma_B} \\ Y_k^{(2)} = \sum_{i=1}^{M_k} x_{ik}^{2\gamma_B} \\ \vdots \\ Y_k^{(N_B)} = \sum_{i=1}^{M_k} x_{ik}^{N_B \gamma_B} \end{array} . \tag{A-2}$$

The relationships described above are based on the conceptual approach of Equation (15). Equation (16) is the exact representation of the error function which the Appendix A approach is based upon.

The calibration parameter vector **P** contains two sets of coefficients, **a** and **b**, as shown in Equation (A-3). The sub-vector **a** is equivalent to the calibration vector **P** in Kasparis *et al.* (2010). This approach results in a method where two sets of independent calibration curves are created: one from the TB data, and one from the optical extinction data. The sub-vectors **a** and **b** are independent and are computed simultaneously. This approach was driven by a desire to define a matrix that could always be inverted to provide the final calibration coefficients. This splitting of calibration coefficients is equivalent to the modified error function described by Equation (15). The model of Equation (8) results in a parameter **P** vector with only components $a$ and $b$. Using the notation of Equation (15), $a = a_A$ and $b = (6/\pi\ a_B)^{(2-\chi)/3}$.

$$\mathbf{P} = \begin{pmatrix} a_1 \\ \vdots \\ a_{N_A} \\ b_1 \\ \vdots \\ b_{N_B} \end{pmatrix} = \mathbf{\Gamma}^{-1} \cdot \mathbf{\Phi} , \tag{A-3}$$

where,

$$\mathbf{\Gamma} = \sum_{k=1}^{M} \left\{ |\mathbf{A}_k\rangle\langle\mathbf{A}_k| + |\mathbf{B}_k\rangle\langle\mathbf{B}_k| \right\} , \tag{A-4}$$

and,

$$\mathbf{\Phi} = \sum_{k=1}^{M} \left( \mathbf{A}_k + \mathbf{B}_k \right) . \tag{A-5}$$

Because of the way the error function is defined, as described in Section 6.2, two simultaneous calibrations result. $D_A$ in Equation (A-6) corresponds to the tipping bucket calibration and $D_B$ in Equation (A-7) to the laser calibration, where $x$ is the impulse amplitude of a drop:

$$D_A = \left( \frac{6}{\pi} \sum_{j=1}^{N_A} a_j\ x^{j\gamma_A} \right)^{1/3} \tag{A-6}$$

$$D_{B_i} = \left( \sum_{j=1}^{N_B} b_j\ x^{j\gamma_B} \right)^{1/(2-\chi)} . \tag{A-7}$$

where $N_A = N_B = 1$ in the special case of the disdrometer response model specified by Equation (8). Forcing the two calibrations to converge (by choice of $\gamma_A$) generates a combined calibration. Even though both solutions are forced to converge as demonstrated in Section 6.2 and 7, a real calibration utilizing Equations (A-6) and (A-7) will not yield exactly equivalent values due to imperfect convergence and numerical noise. Therefore it is useful to merge the two values using an arithmetic average $D = 1000\left(D_A + D_B\right)/2$ or the geometric mean: $D = 1000\left(D_A D_B\right)^{1/2}$ mm.